\documentclass[12pt, letterpaper]{article}

\usepackage{amsmath,amssymb,amsfonts}
\usepackage{algorithm}
\usepackage{algpseudocode}
\usepackage{subcaption}
\usepackage[top=0.70in]{geometry}

\usepackage{graphicx}
\DeclareMathOperator*{\argmax}{arg\,max}

\title{\fontsize{14}{15}\bfseries 
DRo: A data-scarce mechanism to \\ 
revolutionize the performance of Deep \\ Learning based Security Systems}

\author{
Mohit Sewak\\
\texttt{Microsoft R\&D, India}\\
\texttt{mohit.sewak@microsoft.com}
\and
Sanjay K. Sahay, Hemant Rathore\\
\texttt{BITS Pilani, Goa, India}\\
\texttt{\{ssahay, hemantr\}@goa.bits-pilani.ac.in}
}

\date{} 

\begin{document}
\maketitle

\begin{abstract}
Supervised Deep Learning requires plenty of labeled data to converge, and hence perform optimally for task-specific learning. Therefore, we propose a novel mechanism named DRo (for Deep Routing) for data-scarce domains like security. The DRo approach builds upon some of the recent developments in Deep-Clustering. In particular, it exploits the self-augmented training mechanism using synthetically generated local perturbations. DRo not only allays the challenges with sparse-labeled data but also offers many unique advantages. We also developed a system named DRoID that uses the DRo mechanism for enhancing the performance of an existing Malware Detection System that uses (low information features like the) Android \textit{implicit} Intent(s) as the only features. We conduct experiments on DRoID using a popular and standardized Android malware dataset and found that the DRo mechanism could successfully reduce the false-alarms generated by the downstream classifier by $67.9\%$, and also simultaneously boosts its accuracy by $11.3\%$. This is significant not only because the gains achieved are unparalleled but also because the features used were never considered rich-enough to train a classifier on; and hence \textit{no decent performance could ever be reported by any malware classification system till-date using these features in isolation}. Owing to the results achieved, the DRo mechanism claims a dominant position amongst all known systems that aims to enhance the classification performance of deep learning models with sparse-labeled data.
\end{abstract}


\section{Introduction} \label{sec:introduction}

Supervised Deep learning (DL)\cite{sewak-overview-dl} models need plenty of task-specific labelled data to train. Collecting such a large corpus of labelled data can be challenging. Therefore, methods like Weak-Supervision (WS) \cite{snorkel} and Active-Learning (AL) \cite{AL-Survey} use either human-in-loop (HiL) or knowledge-base (KB) supports to generate sizable labelled data, are gaining traction. But the labels generated by these systems are very noisy, and hence not suitable for many critical applications; especially for use in security systems. Also, the HiL and KB supports are more relevant to domains/applications where human intuition/cognizance is considered as the gold-standard.

But generating new labels may not be the only way to uplift the task-specific performance of any DL classification model. One example of this is the Transfer-Learning (TL) \cite{transfer-learning-survey} system. TL is used in many popular vision and transformer based DL models like the BERT \cite{bert}. In TL, most of the trainable parameters of a task-agnostic model are \textit{pre-trained} over a very large corpus of unlabelled data. Subsequently, a few task-specific layers of the model are \textit{fine-tuned} with (relatively) little labelled data. With this unsupervised pre-training approach, TL enables supervised DL models to outperform consistently non-TL models that are trained on similar sized labelled dataset. 

But TL remains a viable option only for domains like language and vision\cite{sewak-cnn}, where acquiring large unlabelled domains/application relevant data is rather inexpensive. For pre-training a Transformer based TL language models like the OpenAI’s GPT-3 \cite{GPT-3} or the Google's BERT \cite{bert}, billions to even trillions of (web) common-crawl documents could be easily and inexpensively made available. But in other domains, collecting such huge data corpus for pre-training may not be possible. Hence, TL could not gain popularity outside of such domains. One such domain where acquiring each data record is very expensive is \textit{security}, especially the area of malware detection. Such applications need malware data, and collection, analysis and validation of each unique malware record requires enormous investments. Hence, there exists no standardized malware dataset of the scale that could be used for pre-training under TL. For such domains, even the techniques like AL and WS are not useful, as for such applications, neither noisy labels are safe, nor could human intuition/cognizance be considered gold-standard.

Therefore, in this paper, we present a novel data-scarce mechanism to achieve the objective of enhancing task-specific (supervised) classification performance under scarce-labelled data scenario. We call this mechanism as \textbf{DRo} (for \textit{\textbf{D}eep-\textbf{Ro}uting}). The DRo mechanism pre-determines which samples could be discriminated well by different downstream classifier in the network, and routes selective samples both for the training of and also for prediction from the different classifiers. Doing so, each model could fit better and faster during training and could be scored efficiently and effectively in production. The DRo mechanism is inspired by some of the very recent research developments in end-to-end (E2E) Deep Clustering (DC) \cite{DEC, imsat}. In particular, DRo exploits the self-augmented training (SAT) mechanism using synthetically generated local perturbations.

The rest of the paper is organized as follows. We cover the background and motivation in section \ref{sec:background}, and some related work in section \ref{sec:related-work}. Then in section \ref{sec:proposed-model-architecture} we cover the mathematical and architectural details of the proposed model and mechanism. Next, we describe the experiments we conducted with the DRoID system in section \ref{sec:experiments}, and conclude the paper in section \ref{sec:conclusion}.

\section{Background \& Motivation}\label{sec:background}

In this section we first cover the mathematical basis of deep-clustering algorithms (subsection \ref{sec:clustering-algorithm} that are modified for the purpose of DRo. To prove the claims of the DRo mechanism, we propose an arduous challenge of making a worth classifier from a non-intuitive, non-enriched features, i.e. the Android implicit-intents. Therefore, next we describe these (subsection \ref{sec:intent}).

\subsection{Deep Clustering: The Mathematical foundation of DRo}\label{sec:clustering-algorithm}\label{sec:dro-mathematics}
Besides being extremely scalable, deep neural networks (DNN) are very flexible in representing complex non-linear decision boundaries. This makes DNN highly suitable for generating discrete representation of multi-dimensional data as required in clustering and hashing.
But DNN are prone to over-fit owing to their high model complexity. In unsupervised learning since the target is not provided the problem is unconstrained, which further exacerbate the over-fit issue. Therefore such models require regularization to attain invariance against minor perturbations atypical to application domain to learn meaningful representations. 

IMSAT \cite{imsat}, one of the recent algorithms in the actively researched area of Deep Clustering currently (refer section \ref{subsec:relatedwork-deep-clustering} for other deep clustering algorithms and their evolution), uses Self Augmented Training (SAT), a mechanism that uses synthetically generated local perturbations from original data samples and rewards their predicted representation to be closer to the representation of the actual data from which the perturbation was generated. It further use a regularization coefficient $\lambda$ to weigh the SAT regularization penalties. Besides using the SAT penalties to provide regularization, IMSAT also use the (Regularized) Information Management (RIM) criteria \cite{NIPS-Regularized-Information-Maximization, NIPS-MI-Phantom-Targets} to ensure that the generated representation are meaningful clusters.

RIM uses a stochastic classifier ($p_{\theta}$) for cluster predictions and rewards its cluster prediction outcomes such that the mutual information amongst the samples in a given cluster is maximized \cite{NIPS-Regularized-Information-Maximization}, while also simultaneously regularizing the classifier complexity.
RIM minimizes the objective as in equation \ref{eq: rim-objective}, where, $X \in \mathbf{X}$ are the data samples, $Y \in \mathbf{Y} \equiv \{0, ..., {K-1}\}$ are the cluster assignments (for $K$ clusters), $\mathcal{R}_\theta$ is the regularization penalty, and $\mathbf{I} (\mathbf{X};\mathbf{Y})$ is MI between data samples and cluster assignments,

\begin{equation}\label{eq: rim-objective}
    \mathcal{R}_\theta - \lambda \mathbf{I} (\mathbf{X};\mathbf{Y})
\end{equation}

Considering multi-dimensional data, the RIM could be expressed as $p_{\theta}(y_1,..., y_m|x)$ by DNN. Assuming that the data dimensions are conditionally independent, the joint probability could be expressed as a product of individual conditional probabilities as in equation \ref{eq: multi-dimension}.

\begin{equation} \label{eq: multi-dimension}
    p_{\theta}(y_1,..., y_m|x) = \prod_{m=1}^{M} {p_{\theta}(y_m|x)}
\end{equation}
If $T : \mathbf{X} \longrightarrow \mathbf{X}$ is the transformation function that generates the local perturbations (data augmentation) to ensure invariance, the SAT regularization penalty for the predictions $p_{\hat\theta}(y_m|x)$, generated by a network with parameter $\hat \theta$, for the original data point $x$ augmented with perturbations $T(x)$ could be expressed as equation \ref{eq: sat}.
\begin{equation} \label{eq: sat}
    \begin{aligned}
     \mathcal{R}_{SAT}(\theta; x, T(x)) = \\
      - \sum_{m=1}^{M} \sum_{y_m=0}^{V_m-1}  p_{\hat\theta}(y_m|x)\log p_{\theta}(y_m|T(x))   
    \end{aligned}
\end{equation}

Similarly SAT penalty for the entire dataset $\mathbf{X} (x \in \mathbf{X})$ is the average of individual $\mathcal{R}_{SAT}(\theta; x, T(x))$, and is given as in equation \ref{eq: rsat-averaging}.

\begin{equation} \label{eq: rsat-averaging}
     \mathcal{R}_{SAT}(\theta;T) =
      \frac{1}{N} \sum_{n=1}^N \mathcal{R}_{SAT}(\theta; x_n, T(x_n))
\end{equation}
If $r$ is a small perturbation that does not alter the meaning of the data point for the given context, then in the simplest form, an illustrative augmentation function could be expressed as $T(x)=x+r$. Regularization against such small, local perturbations, $r$, enforces local-invariance in the data-representations, and hence ensures that the cluster-separation boundaries lie in the region where data-distribution-density is relatively low. Such cluster-separation-boundaries adhere to the low-density-separation principle and are hence preferable.
IMSAT use a more sophisticated data augmentation method, called the Virtual Adversarial Training (VAT), in which the local-perturbation $r$ is computed in an adversarial setting \cite{Distributional-Smoothing-Virtual-Adversarial}. This is given in equation \ref{eq: r-vat}.
\begin{equation}\label{eq: r-vat}
    r=\argmax_{r\prime}{R_{SAT}(\theta;x,x+r\prime); \vert\vert r\prime \vert\vert_2 \leq \epsilon}
\end{equation}
If $H(Y)$ is the marginal-entropy (ME) of the data (as expressed in equation \ref{eq: marginal-entropy}), and $H(Y|X)$ is the conditional-entropy (CE) of the cluster-assignments of the data given the features (as expressed in \ref{eq: conditional-entropy}), the MI gain due to \textit{clustering} could expressed as a difference between ME and CE \cite{book-information-theory}. Therefore the total loss $Loss_{total}$ as a function of loss due to regularization $Loss_{R_{SAT}}$ and loss due to clustering (negative information gain), as moderated by the coefficient $\lambda$, could be given as in equation  \ref{eq: total-loss}.
\begin{equation}\label{eq: total-loss}
    Loss_{total} = Loss_{R_{SAT}}-\lambda{H(Y)-H(Y|X)}
\end{equation}
\begin{equation}\label{eq: marginal-entropy}
    H(Y) \equiv h(p_\theta(y)) = h(\sum_{i=1}^N p_\theta(y|x))
\end{equation}
\begin{equation}\label{eq: conditional-entropy}
    H(Y|X) \equiv \frac{1}{N}\sum_{i=1}^N h(p_\theta(y|x))
\end{equation}
where $h(p(y)) \equiv -\sum_{y\prime} p(y\prime) log p(y\prime)$ is the entropy function.
Since increasing the ME could enforce the cluster sizes to be uniform, while decreasing the CE could enforce unambiguous cluster assignments \cite{NIPS-MI-Phantom-Targets}, we add an extra parameter $\mu \in \mathbb{R}$ in DRo to further tune the CE in the \textit{`DRoID' or `Vault'} mode 
(figure \ref{fig:ids-vault-architecture})
settings. 
So the DRo's loss could be calculated as in equation \ref{eq: mu}.
\begin{equation}\label{eq: mu}
    Loss_{clustering} = - (H(Y) - \mu H(Y|X))
\end{equation}

\subsection{Android \textit{Implicit} Intent}\label{sec:intent}
Android OS is a sophisticated eco-system of multiple apps that in aggregation work to provide a great user-experience and enhance productivity of both the end-consumer and the app-developers. Because of a mechanism called Intent, the app-developers could modularise their code into components, where each component could be coded to perform a few closely tied activities. The app could call other components to perform any ancillary functionality that the developer does not intend to code in a particular component. If the required component is pre-defined by the developer these could be declared as explicit-intent within the code, and could be sent to the Android system for execution during run-time. 

Alternatively, if these components are not pre-defined by the developer, these are sent to the Android system as a general intent for which the Androids systems need to first discover all apps that have a component that could effectively fulfill such an intent. Any app that can handle any type of implicit-intent, states the activities that it could fulfil in an implicit-intent filter in its readily discoverable implicit-intent \textit{xml} file. There could be multiple applications that could declare the same activity in their implicit-intent filter. In such a scenario, the Android system creates an interrupt for the user to select the best desired app to fulfil that intent if any app has not been set as default by the user, or the system. A very intuitive example of this is browsing activity. When a user clicks an URL in an app that cannot handle browsing activity, the browsing intent is sent to the Android system, which in turns finds the apps that could open the URL, e,g, browser. If there is over one browser installed in the system, then Android creates a pop-up for the user to choose a specific browser.

Unlike Android-permissions, which are under the control of a malicious app developer to request and to execute, the implicit-intents could only be declared by the app-developer. Which application could call them, and if at all these are called, and if called, if the user selects the specific app to execute it are not in the app developers control and hence counter-intuitive to be used by a malware developer. Hence the implicit-intent, though are easy-to-extract, may not offer very rich features for designing a malware classifier. We cover more on this aspect in section \ref{sec:related-work}.

\section{Related Work} \label{sec:related-work}

Since DRo is mathematically inspired from developments in Deep Clustering, and we validate its claims by implementing it in an online mobile IDS that use low-information static features (like the Android Intent), so we present the related work corresponding to both of these aspects and accordingly split this section into two sub-sections as follows.

\subsection{Related Work in the area of Deep Clustering} \label{subsec:relatedwork-deep-clustering}
In the last 3 to 4 years, E2E DL for generating a discrete representation of data (clustering and hashing) has gained a lot of traction in research fraternity and different algorithms have been proposed that have been proposed to this end. Such discrete representation could be for clustering or hashing. When Deep Learning is used for clustering, we call it Deep Clustering (DC). Many of the conventional and popular \textit{representative} clustering and hashing algorithms like Gaussian Mixture Modeling (GMM) and K-Means are incapable of modelling non-linear boundary separation between clusters. Others like the \textit{kernel} \cite{NIPS-Maximum-Margin-Clustering, NIPS-Binary-Reconstructive-Embeddings} and \textit{spectral} \cite{NIPS-Spectral-Hashing} clustering-based techniques though can model arbitrary cluster boundaries, but are not scalable to large datasets. Here is where DC shines as it model non-linear and complex separation boundaries and is extremely scalable and flexible. Many of the recent algorithms in this field are also capable of producing data embeddings along with the cluster representations.

Deep Embedding for Clustering (DEC) \cite{DEC} was the first DC algorithm that could simultaneous generate embedding and clustering representations. DEC did not use generative or variational augmentations of data for regularization. Other similar approaches for generating adversarial robustness were proposed by \cite{variational-deep-embedding}\cite{sewak2020doom}\cite{rathore2020robust}. In these approaches, the data is modelled by a \textit{generation} process, and they use GMM as prior distributions for deep generative models. This approach is atypical to data generation using variational AE. Leen \cite{DL-Regularization-Invariant-Learning} proved that applying data augmentation to a DL classifier gives similar effect as applying regularization to the original cost function of a conventional machine learning (ML) classifier. Miyato et al.\cite{Distributional-Smoothing-Virtual-Adversarial}, and Sajjadi et al.\cite{NIPS-Perturbations-Regularization} successfully adapted this approach to semi-supervised DL algorithms and demonstrated successful results. SAT is inspired by this approach of regularization. Dosovitskiy et al.\cite{NIPS-Discriminative-Unsupervised-Feature-Learning} proposed to use data augmentation to model the invariance of learned representations in even unsupervised algorithms like clustering. IMSAT takes similar approach, but applies invariance directly to the learned representation instead of applying it on the surrogate classes, and directly learns discrete representations (clusters), instead of learning continuous representations that are later converted to discrete class-predictions or cluster-assignments.

\subsection{Related work on Android IDS and Android Intent features} \label{subsec:relatedwork-intent}

Most personal edge-devices, like smartphones, have two things in common, Android OS and a mandate for battery-efficient endpoint protection platform (EPP). Therefore, besides signature-matching, these devices essentially rely on simple static features for malware detection. Thus, Android-permissions is an actively researched feature for Android EPP. 

One work \cite{PScout} on analysis of this feature shows that variance in specification vs. usage of the documented Android APIs. This inspired many work that use these features for Android EPP; one such example is Lindorfer et al. ‘s ANDRUBIS \cite{ANDRUBIS}. In this work, the researchers discovered that the malicious vs. non-malicious applications have slightly distinct patterns of using permissions, and the malicious apps requests for many extra permissions than their non-malicious counterparts. Other researchers have found using Android-permission for making machine-learning classifiers either in isolation (\cite{li-permission}\cite{rathore2020identification}), others have used it in combination with other, computationally more expensive features like opcodes (\cite{rathore2018android}\cite{rathore2020detection}, \cite{Intent-Permission-Idrees}), or multiple other features like API-calls and Android-components (\cite{DroidSieve}) and achieved better performance. 

 Opcodes though are static features, but are more-computationally expensive to extract. API-calls, on the other hand,a are dynamic features, and are best extracted offline via a cloud based service. Android-implicit-intent, though co-exists with Android-permissions in the manifest but still few researchers have used it. One work that uses Android-intent, although with other complex dynamic and static features, is by Wu et al. \cite{DroidMat}. Another work by Arp et al. \cite{arp-drebin}, besides using most of the above covered complex features, also leveraged network address, to maximize on the feature-set and hence the result. The reasoning for using Android-intent here seems to leverage its efficiency to make an efficient production-ready system but to maximize on the theoretical feature availability. The only system that uses only Android-intent is by Ali et al. (\cite{AndroDialysis}). But the work was more inspired to use Android-intent as a theme, therefore they also extracted the Android-explicit-intents, which is almost similarly more involved as other complex features discussed in other works.
 
 Through Android-implicit-intent is as easy to extract as Android-permissions, it is hard to find any work that uses these as exclusive features for designing an efficient and effective edge or cloud EPP. But as illustrated in a recent work by Sewak et al. (\cite{sewak-pimrc-deepintent}), this is, however, not completely counter-intuitive. Sewak et al. used both an analysis approach and a DL classifier approach to investigate the usefulness on Android-implicit-intent features for malware-detection. They acknowledged that using the features in isolation was challenging and hence they intent to create a baseline for these features (instead of using these in isolation in a production-ready system). They used very sophisticated e2e DL architectures to do malware classification using Android-implicit-intent features only. Though the intention of their work was not to use the Android-implicit-intent features in isolation for an EPP, but the idea is not without its merit given the efficiency requirements for edge-device. Therefore, we believe this quest provides the best scenario to test the claims of the DRo system.

\section{Proposed Model \& Architecture}\label{sec:proposed-model-architecture}

In this section we cover an illustrative architecture for implementing a DRo based system like DRoID in subsection \ref{sec:dro-architecture}. Next we cover the advantages and limitation of the DRo system based on this architecture in subsection \ref{sec:dro-advantages}, and the process pre-processing the dataset for DRoID in subsection \ref{sec:data-preprocessing}.

\subsection{Architecture of a DRo based systems}\label{sec:dro-architecture}
\begin{figure*}[!htbp]
    \centering
    \includegraphics[width=\textwidth]{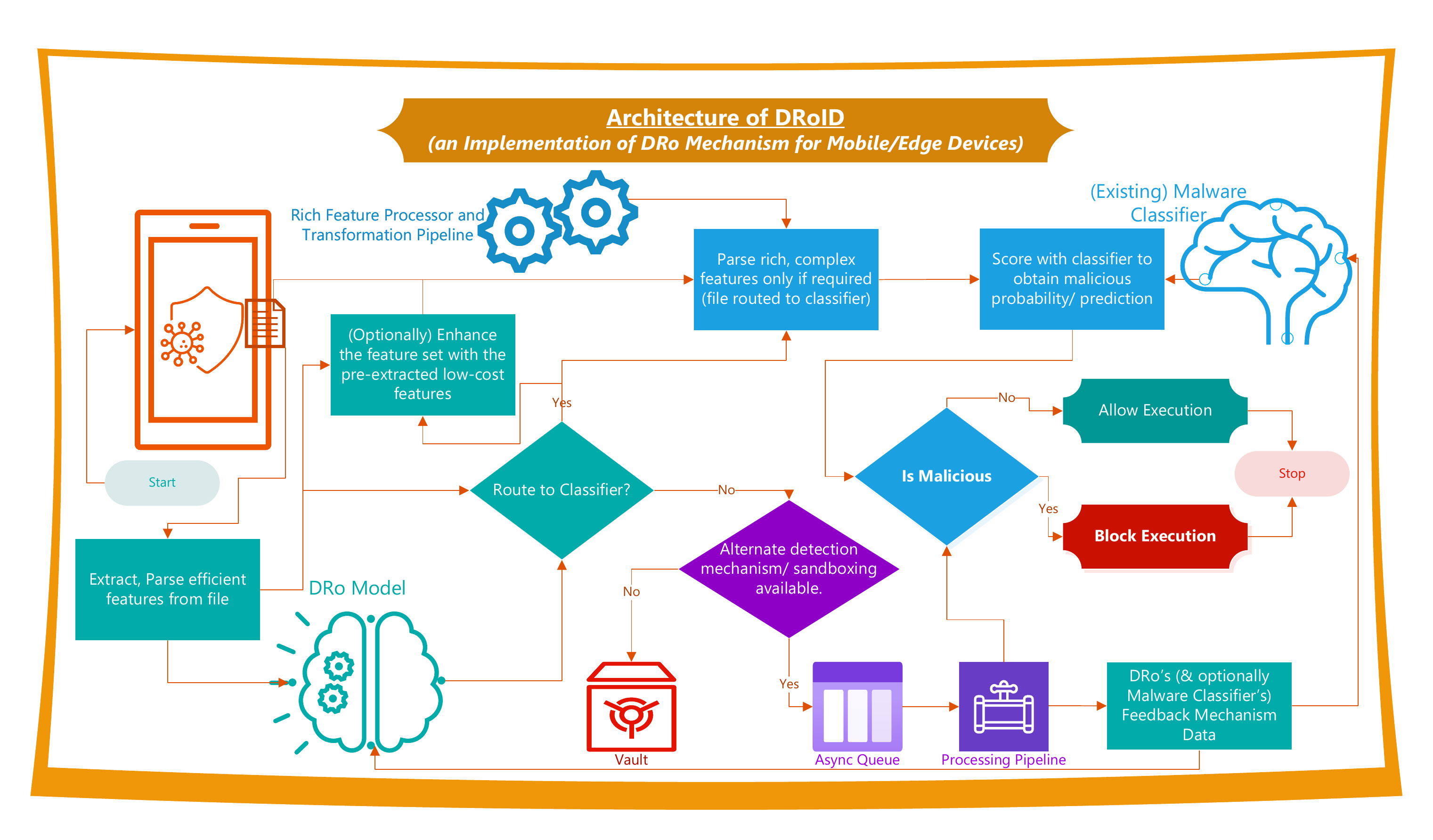}
    \caption{
    Architecture of a DRo based Malware Detection, Protection \& Management System for Mobile \& Embedded devices.}
    \label{fig:ids-vault-architecture}
\end{figure*}

Though, DRo could be used in any field, one domain where it may become indispensable is security, especially for enhancing the performance of Malware/Intrusion Detection Systems (MDS/IDS). The recent research trends in IDS indicate the shift towards using data hungry DL models \cite{rathore2018malware}, where the application of DRo can be revolutionary. Modern IDS, especially the ones used in mission-critical applications, often use layered analysis. In such layered IDS, the online classifier works on efficient, and cost-effective static features. Such IDS may also have a secondary, and more costly detection-mechanism in the form of a dynamic analysis layer. In such layered/hybrid IDS, costlier dynamic analysis is applied only to the samples that are not optimally covered (not classified, or classified with insufficient confidence) by the online classifier. The DRo mechanism when used to augment such an IDS, could automatically route the incoming file samples either to the online classifier or to a \textit{Vault} which could serve as a queue mechanism for the dynamic analysis scheduler and processor.
We call such an implementation of DRo as \textit{DRoID} (\textbf{DRo} for \textbf{I}ntrusion \textbf{D}etection). The architecture for such an atypical DRoID system is shown in figure \ref{fig:ids-vault-architecture}. The algorithm algorithm/pseudo-code for training a DRo based system is as provided in algorithm \ref{alg:experimental-algo})

\begin{algorithm}[hbp!]
\footnotesize
	\caption{Algorithm for Training a DRo based system}\label{alg:experimental-algo}
	\textbf{Input:} \\
	$\mathbf{\lambda}$:  trade-off parameter between Re-constitution error and Clustering Loss \\
	$\mathbf{\mu}$: multiplier for Conditional Entropy Loss \\
	$\mathbf{e}$: Number of training epochs \\
	$\mathbf{T}$: set of training sample \\
	$\mathbf{V}$: set of validation sample \\
	$\mathbf{C}$: set of malware detection models \\
	\textbf{Function} : \\
	$\mathbf{C_P}$: prediction function in classification models trained on P\\
	$\mathbf{F_m}$: feature vector modification function \\
	$\mathbf{R}$: routing clustering function \\	
	\textbf{Output} : ${Sample (s_g, n_c, c)}$\\
	$\mathbf{N_c}$: route-cluster assignment of the candidate sample\\
	$\mathbf{L_{v}}$: reconstitution loss \\
	$\mathbf{L_{routing}}$: routing/ clustering loss\\
	$\mathbf{L_{total}}$: total loss\\
	$\mathbf{E_m}$: Embedded dimension of $E_{layers}-1$ of Encoder\\
	$\mathbf{Lift_a}$: Router accuracy lift for r=1\\
	\begin{algorithmic}[1]
	    \State $a_{benchmark} \leftarrow C_T(V)$ 
	    \State $N_{routes} \leftarrow 2$ \Comment{r<-0 indicates ambiguity and r<-1 indicated dis-ambiguity}
	    \For{each epoch in [1, e]}
		    \For {each batch b $\mathbf{\in}$ T}
                    \State $L_{mi} \leftarrow E(Y) - \mu \times {E(Y | X)}$
                    \State $L_{total} \leftarrow L_{v} + \lambda \times L_{mi}$
                    \State $W_t = W_{t-a} - \alpha \times (\delta L_{total})$
          	    \EndFor
    		\EndFor
    	\For{each batch b $\mathbf{\in}$ T}
    	    \State $r \leftarrow R^t(b)$
    	    \State ${f = F_m^t(b)}$
    	       \If {${(r == 1)}$}
            		    \State ${C_T = C(f,b)}$
        		\Else
        		    \State route sample to alternative-analysis/ vault
        		\EndIf
        	\State ${a_b \leftarrow N(C_T(V_b)==Y(V_b))/N(b)}$ 
    	\EndFor
    	\State ${a_{deeprouter} \leftarrow average(a_{b1},...,a_{bn})}$ 
    	\State ${Lift_a \leftarrow a_{deeprouter} - a_{benchmark}}$
	\end{algorithmic} 
\end{algorithm}
\raggedbottom

To put DRo/DRoIDs claims to test, we take an extreme example. We build a mobile IDS using the DRoID architecture, and in it we use the Android \textit{Implicit Intent} features \ref{sec:intent} alone. These features are not considered very insightful for developing an IDS \cite{sewak-pimrc-deepintent}, and hence no IDS exist (beyond benchmarks) that use these features in isolation \cite{sewak-pimrc-deepintent}. From the experiments conducted on DRoID using a popular and standard Android malware (\textit{Drebin}) dataset \cite{arp-drebin}, we found that the DRo mechanism could successfully reduce the \textbf{false-alarms} generated by the downstream classifier by up to $\mathbf{67.9\%}$, while also simultaneously boosting its \textbf{accuracy} by $\mathbf{11.3\%}$. To the best of our knowledge, this quantum of enhancement to an existing IDS classifier (especially without re-training the classifier or using additional auxiliary data) is unheard of in any available public literature. Hence, this extreme and challenging illustration provides utmost credibility to the claims of the DRoID system and the DRo mechanism.

\subsection{Advantages \& Limitations of DRo} \label{sec:dro-advantages}

Unlike TL, DRo does not replace/substitute an existing classifier, but could alternatively be used to augment it. This is a very significant and revolutionary shift, as because of this, DRo could even be retrofitted to any existing classification system, and does not mandates a redesign and retraining of the system. There are several other advantages that DRo offers over TL, these are enumerated next.

\begin{enumerate}
    \item Where TL is incompatible with existing DL/ML classifiers, and mandates to replace them for any incremental performance improvements, DRo aims to augment an existing classification system, and does not mandate any changes to an existing classifier, not even re-training it.
    \item Unlike TL, DRo can be developed and re-trained independent of the task-specific model.
    \item DRo is based on deep-clustering architectures which have a much smaller model foot-print and hence are economical to train than their TL counterparts.
    \item DRo can also work on much simpler features than that used by the classifier (refer section \ref{sec:intent}). Coupled with smaller size, DRo based systems offer easy and efficient online deployment opportunities, without mandating distillation. \label{dro-adv:simpler-features}
    \item DRo does not require (large) corpus of un-labelled data. It can even work on the same task-specific labelled dataset as the task-specific model and yet improve its performance. 
    \item DRo can be retrofitted to any existing task-specific model without mandating it to be re-trained or even (re-)fine-tuned.
    \item DRo offers on-demand coupling, where if required the feedback from the task-specific model can be served to improve DRo (fig: \ref{fig:ids-vault-architecture}).
    \item The on-demand coupling also work in the reverse direction, where the output from DRo can also be optionally served as input to the downstream model (fig: \ref{fig:ids-vault-architecture}) \label{dro-adv:reverse-on-demand-coupling}.
    \item DRo can uplift a model's performance (Table:\ref{table:experiments-lam-acc}), and also influence the accuracy-FPR balance of a downstream classifier without re-training the model or altering any of its parameters.
\end{enumerate}

These benefits come at a cost of reduced coverage for classification for a system augmented with DRo. This aspect may look like a shortcoming prima-facie, but is actually an added advantage. For systems which need to capture noisy data so as to augment the classification system, this aspect is an added advantage as this process is automated by DRo. For systems that favors, coverage over false-alarms, the reduced coverage records could still be used with \textbf{known} classification limitations for those records, a feature that any standard classifier cannot offer by default.

\subsection{Dataset and its Pre-Processing} \label{sec:data}\label{sec:data-preprocessing}
\begin{figure}[htb!]
    \centering
    \includegraphics[height=\textheight]{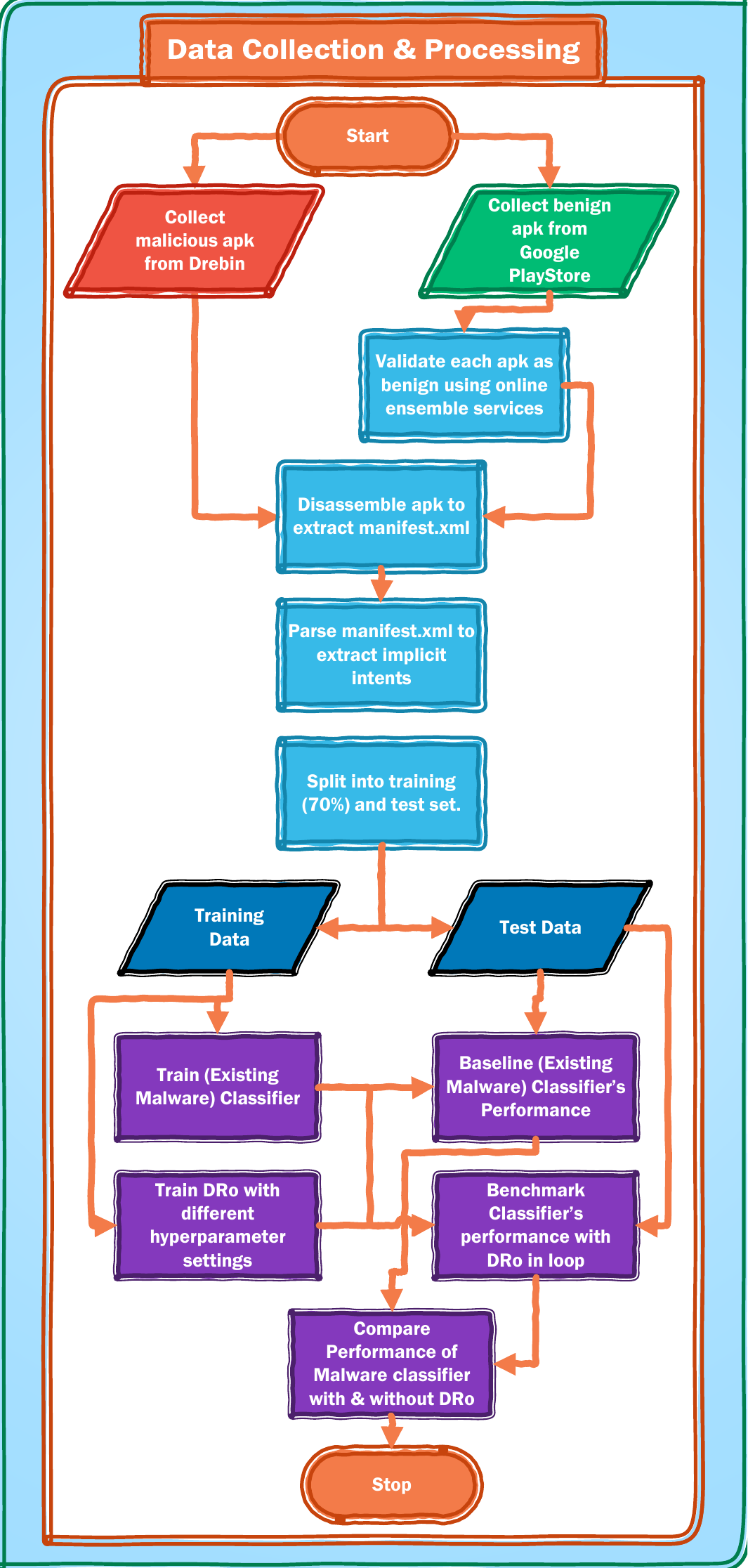}
    \caption{Data Collection and Processing}
    \label{fig:data-process}
\end{figure}

We use a very popular standardized Android malware dataset \textit{`Drebin’} dataset\cite{arp-drebin} for our experiments. This is an old and had been an actively researched dataset in the past, and hence we assume that maximum performance that could be derived from this dataset is available in the literature. As opposed to producing benchmarks on an under-explored dataset, we prefer using such adequately explored dataset to ensure that the improvement achieved rightly attribute credibility to DRo mechanism and not application of a different classifier-parameter setting or any different feature-extraction technique. The dataset contains  $\approx 5,000$ malicious apk samples across $20$ different families. Therefore, to balance this positive dataset, we collected similar sized negative dataset, and verified that each sample of the negative dataset in non-malicious using the ensemble classification from the \textit{VirusTotal} utility. From each of the collected sample, we extracted their respective `manifest.xml’ (manifest) file using the open-source \textit{APKTOOL} utility. Next, we parsed the manifest using a custom-utility to extract all implicit-intents as declared under the intent-filter section in the manifest file and mapped them to one of $273$ unique Android-implicit-intent types. Next, using these raw features we trained a series of benchmark classifiers without the DRo. The same dataset was also used for experiments with the best benchmark classifier, augmented with DRo for comparison. This entire process is summarized in figure \ref{fig:data-process}.

\section{Experiments and Results} \label{sec:experiments}\label{sec:results}
To establish the improvement that could be attributed to the DRo mechanism, we conducted experiments in 2 series.
In the first series of experiment, we benchmark the performance of an atypical (\textit{`existing'}) IDS classifiers that have provided leading results in the work covered in section \ref{subsec:relatedwork-intent}. In the next section, we repeat the experiments with DRo enabled DRoID on the same dataset and the `same' classifier. The baseline models whose performance were bench-marked in the first series are shown in table \ref{table:experiments-benchmark}. In this benchmark the Random Forest (RF) provided the best results. To eliminate chances of any bias in favor of DRo, we use this `best' bench-marked classifier as the baseline performance, and hence we will be using this configuration as a downstream classifier to compare the different configuration of DRoID. As indicated in figures \ref{fig:ids-vault-architecture}, and advantage \ref{dro-adv:reverse-on-demand-coupling}, the features generated from the DRo could be shared with the downstream classifier, but to avoid any positive bias in favor of DRo arising from the inclusion of these richer features, we use the best benchmark-ed classifier (RF) \textit{as-is} on raw (sparse) features only. Also as per figure \ref{fig:data-process} several experiments were carried with different configuration (and 2 setups) of the DRoID. Each of these experiments followed the algorithm \ref{alg:experimental-algo}.

\begin{table}[hbp!]
\small
\begin{center}
\begin{tabular}{|c|c|c|c|}
\hline
\textbf{B.ID.} & \textbf{Algorithm} & \textbf{Accuracy} & \textbf{FPR} \\ 
\hline
\textbf{B1} & \textbf{Random Forest} & \textbf{0.777} & \textbf{0.106} \\
B2 & AdaBoost & 0.761 & 0.127 \\ 
B3 & 1-L MLP-DNN & 0.741 & 0.116 \\ 
B4 & 2-L MLP-DNN & 0.745 & 0.138 \\ 
B5 & 3-L MLP-DNN & 0.752 & 0.111 \\ 
B6 & 4-L MLP-DNN & 0.772 & 0.11 \\
\hline
\end{tabular}

\caption{Baseline classifier performance without DRo}
\label{table:experiments-benchmark}
\end{center}
\end{table}

Table \ref{table:experiments-lam-losses} and \ref{table:experiments-mu-losses} are conducted with the same classifier as assisted with the DRoID system (fig: \ref{fig:ids-vault-architecture}). In these configurations DRoID is fine-tuned mathematically to alter its available adjustable parameters (refer section \ref{sec:dro-mathematics}). The tables \ref{table:experiments-lam-losses} indicates the effect of varying the regularization parameter $\lambda$. The results indicates that keeping the value pf $\mu$ constant (at 4) a good balance between lower $Loss_{total}$, higher $H(Y|X)$ (or lowest $-H(Y|X)$) and lower $R_{SAT}$ could be reached at $\lambda=0.4$. At any value of $\lambda > 0.4$ CE is almost $0$. Though values of $\lambda < 0.4$ could also be considered.

\begin{table}[hb!]
\begin{center}

\begin{tabular}{|c|c|c|c|c|c|}
\hline
\textbf{\begin{tabular}[c]{@{}c@{}}Conf. \\ ID\end{tabular}} & $\mathbf{\lambda}$ & $\mathbf{\mu}$ & $\mathbf{L_{total}}$ & $\mathbf{H(Y|X)}$ & $\mathbf{R_{SAT}}$ \\
\hline
1 & 0 & 4 & 0.132 & 0.152 & 0.132 \\
2 & 0.1 & 4 & -0.228 & 0.002 & 0.049 \\
3 & 0.2 & 4 & -0.482 & 0.001 & 0.071 \\
4 & 0.3 & 4 & -0.755 & 0.001 & 0.074 \\
5 & \textbf{0.4} & 4 & \textbf{-1.018} & \textbf{0.001} & \textbf{0.086} \\
6 & 0.5 & 4 & -1.318 & 0 & 0.066 \\
7 & 0.7 & 4 & -1.902 & 0 & 0.036 \\
8 & 0.8 & 4 & -2.161 & 0 & 0.054 \\
9 & 0.9 & 4 & -2.449 & 0 & 0.043 \\
10 & 1 & 4 & -2.713 & 0 & 0.055 \\
\hline
\end{tabular}

\caption{Configurations and Losses with different values of $\lambda$}
\label{table:experiments-lam-losses}
\end{center}
\end{table}

To validate these findings and identify a specific value of $\lambda$, we conduct experiments with layered DRoID with hybrid online-classifier and dynamic-analysis.
Table \ref{table:experiments-lam-acc} shows these results. Here route $C1$ belongs to the online downstream (baseline) classifier. Route $C0$ indicated to the Vault in \textit{Vault} mode, in which we have put another similar classifier to mimic the dynamic-analysis mode to assess the dis-ambiguity in both the routes. The classifiers used for the assessment (RF) and the associated features (raw, sparse, \textit{implicit} Intent(s)) are the same as indicated in the benchmark setup.

These results validate that the regularized \textit{configuration-5}, with \textit{$\lambda=0.4$} provides the best result for the constant $\mu$. In this configuration, $51\%$ of the data sample flows into the route where the (baseline) classifier was, and this data could be unambiguously classified with much better accuracy of $86.5\%$ as compared to $77.7\%$ (boost of $\mathbf{11.3\%}$) and even better FPR of $0.034$ as compared to $0.106$ (reduction in false alarms by $\mathbf{77.7\%}$) of the best baseline/benchmark classification. The results also indicate that the samples in the $C0$ route could not be classified decently, hence validating that these were ambiguous and hence correctly sent to the vault for dynamic-analysis in a layered IDS.

\begin{table}[hb!]
\begin{center}
\begin{tabular}{|c|c|c|c|c|c|}
\hline
\textbf{\begin{tabular}[c]{@{}c@{}}Conf. \\ ID\end{tabular}} & \textbf{\begin{tabular}[c]{@{}c@{}}Ratio \\ C0:N\end{tabular}} & 
\textbf{\begin{tabular}[c]{@{}c@{}}C0 \\ Acc.\end{tabular}} &
\textbf{\begin{tabular}[c]{@{}c@{}}C0 \\ FPR\end{tabular}} &
\textbf{\begin{tabular}[c]{@{}c@{}}C1 \\ Acc.\end{tabular}} &
\textbf{\begin{tabular}[c]{@{}c@{}}C1 \\ FPR\end{tabular}} \\
\hline
1 & 0.32 & 0.849 & 0.176 & 0.722 & 0.176 \\
2 & 0.5 & 0.689 & 0.032 & 0.856 & 0.032 \\
3 & 0.5 & 0.687 & 0.033 & 0.857 & 0.033 \\
4 & \textbf{0.49} & 0.685 & 0.032 & \textbf{0.861} & \textbf{0.032} \\
5 & \textbf{0.49} & 0.68 & 0.034 & \textbf{0.865} & \textbf{0.034} \\
6 & 0.54 & 0.406 & 0.838 & 0.858 & 0.838 \\
7 & 0.54 & 0.406 & 0.838 & 0.859 & 0.838 \\
8 & 0.54 & 0.405 & 0.837 & 0.857 & 0.837 \\
9 & 0.54 & 0.406 & 0.837 & 0.861 & 0.837 \\
10 & 0.54 & 0.406 & 0.837 & 0.858 & 0.837 \\
\hline
\end{tabular}

\caption{Coverage ($1 - {Ratio\frac{C0}{N}}$) and Classifier Performance (C1 Acc., C1 FPR) with different values of $\lambda$ in a layered/hybrid IDS}
\label{table:experiments-lam-acc}
\end{center}
\end{table}

Next, we freeze the value of the parameter $lambda$ at $lambda=0.4$ as obtained from the earlier results and altered the value of parameter $\mu$. The parameter $\mu$ is altered to provide a varying degree of trade-off between the accuracy and false-alarms (FPR) of the classifier in route $C1$. Table \ref{table:experiments-mu-losses} shows these experiments and indicates that even for a constant value of CE $H(Y|X)$ the value of $L_{total}$ increasing with higher $\mu$ multipliers for the CE. The implications of these changes could be seen when similar configurations are viewed in \textit{Layered/Hybrid-IDS} mode as shown in table \ref{table:experiments-mu-acc}. 

\begin{table}[htb]
\begin{center}

\begin{tabular}{|c|c|c|c|c|c|}
\hline
\textbf{\begin{tabular}[c]{@{}c@{}}Conf. \\ ID\end{tabular}} & $\mathbf{\lambda}$ & $\mathbf{\mu}$ & $\mathbf{L_{total}}$ & $\mathbf{H(Y|X)}$ & $\mathbf{R_{SAT}}$ \\
\hline
11 & 0.4 & 1 & -0.205 & 0.001 & 0.071 \\
12 & 0.4 & 3 & -0.746 & 0.001 & 0.083 \\
13 & 0.4 & 5 & -1.285 & 0.001 & 0.095 \\
14 & 0.4 & 7 & -1.846 & 0.001 & 0.089 \\
15 & 0.4 & 9 & -2.415 & 0.001 & 0.077 \\
\hline
\end{tabular}

\caption{Configurations and Losses with different values of $\mu$}
\label{table:experiments-mu-losses}
\end{center}
\end{table}

This result demonstrates that as $\mu$ is increased initially we could strike a trade-off between better accuracy at the cost of greater FPR. Increasing $\mu$ even further deteriorates both accuracy and FPR. Configuration-11 provides slightly lower accuracy while decreasing false alarms. Whereas, Configuration-11 raises the FPR slight to improve upon the accuracy. Anyone of the configurations could be selected depending upon the domain requirement.

\begin{table}[hb!]
\begin{center}
\begin{tabular}{|c|c|c|c|c|c|}
\hline
\textbf{\begin{tabular}[c]{@{}c@{}}Conf. \\ ID\end{tabular}} & \textbf{\begin{tabular}[c]{@{}c@{}}Ratio \\ C0:N\end{tabular}} & 
\textbf{\begin{tabular}[c]{@{}c@{}}C0 \\ Acc.\end{tabular}} &
\textbf{\begin{tabular}[c]{@{}c@{}}C0 \\ FPR\end{tabular}} &
\textbf{\begin{tabular}[c]{@{}c@{}}C1 \\ Acc.\end{tabular}} &
\textbf{\begin{tabular}[c]{@{}c@{}}C1 \\ FPR\end{tabular}} \\
\hline
\hline
11 & 0.48 & 0.684 & 0.017 & \textbf{0.859} & \textbf{0.017} \\
12 & 0.49 & 0.686 & 0.032 & \textbf{0.864} & \textbf{0.032} \\
13 & 0.49 & 0.683 & 0.034 & 0.855 & 0.034 \\
14 & 0.49 & 0.683 & 0.034 & 0.857 & 0.034 \\
15 & 0.5 & 0.689 & 0.034 & 0.852 & 0.034 \\
\hline
\end{tabular}

\caption{Coverage ($1 - {Ratio\frac{C0}{N}}$) and Classifier Performance (C1 Acc., C1 FPR) with different values of $\mu$ in a layered/hybrid IDS}
\label{table:experiments-mu-acc}
\end{center}
\end{table}

\section{Conclusion} \label{sec:conclusion}
In this paper we proposed DRo, a novel data-scarce alternative to TL, WS, and AL for enhancing performance of supervised DL models under scarce-label scenario. DRo neither requires a large corpus of unsupervised data that is required for pre-training in TL; nor does it require human-judgement or generates noisy-labels as with WS or AL. Further, to implement DRo no changes to or re-training of the existing classifier is required and it is also more efficient to train and infer a classifier with DRo. We benchmarked DRo’s capability and established that it could not only reduce the false-alarms generated by a DL classifier by $67.9\%$, but can also achieve the contradictory goal of simultaneously uplifting its accuracy by $11.3\%$. Not only such performance gains are unparalleled by any other system, but it is also capable of training a classifier even under abysmal feature scenarios. The results suggest that the novel DRo mechanism is a very able and feature rich alternative to TL, WS, and AL mechanisms for data-scarce domains and applications.

\bibliography{main}
\bibliographystyle{unsrt}

\end{document}